
\magnification=1200
\hsize 15true cm \hoffset=0.5true cm
\vsize 23true cm
\baselineskip=15pt

\nopagenumbers

\font\grande=cmr10 scaled \magstep4
\font\medio=cmr10 scaled \magstep2
\outer\def\beginsection#1\par{\medbreak\bigskip
      \message{#1}\leftline{\bf#1}\nobreak\medskip\vskip-\parskip
      \noindent}

\def \me {\buildrel <\over \sim}
\def \Me {\buildrel >\over \sim}
\def \pa {\partial}

\def \Da {\Delta}

\def \a {\alpha}

\def \Ga {\Gamma}
\def \ga {\gamma}

\def \da {\delta}

\def \noi {\noindent}

\def\sqr#1#2{{\vcenter{\hrule height.#2pt\hbox{\vrule width.#2pt
height#1pt \kern#1pt\vrule width.#2pt}\hrule height.#2pt}}}

\def\lsim{\mathrel{\rlap{\lower4pt\hbox{\hskip1pt$\sim$}}
    \raise1pt\hbox{$<$}}}         
\def\gsim{\mathrel{\rlap{\lower4pt\hbox{\hskip1pt$\sim$}}
    \raise1pt\hbox{$>$}}}         

\line{\hfil DFTT-03/94}
\line{\hfil January 1994}
\vskip 2 cm

\centerline {\grande  Relaxed Bounds on the Dilaton Mass}
\vskip 1 true cm
\centerline{\grande In a String Cosmology Scenario}
\vskip 2true cm
\centerline{M.Gasperini}
\centerline{\it Dipartimento di Fisica Teorica dell'Universit\`a,}
\centerline{\it Via P.Giuria 1, 10125 Torino, Italy,}
\centerline{\it Istituto Nazionale di Fisica Nucleare, Sezione di Torino}
\vskip 2 cm
\centerline{\medio Abstract}

\noindent
We discuss bounds on the dilaton mass, following from the cosmological
amplification of the quantum fluctuations of the dilaton background, under the
assumption that such fluctuations are dominant with respect to the classical
background oscillations. We show that if the fluctuation spectrum grows with
the frequency the bounds are relaxed with respect to the more conventional case
of a flat or decreasing spectrum. As a consequence, the allowed range of masses
may become compatible with models of supersymmetry breaking, and with a
universe presently dominated by a relic background of dilaton dark matter.

\vfill\eject
\footline={\hss\rm\folio\hss}
\pageno=1

\centerline{\bf Relaxed
bounds on the dilaton mass in a string cosmology scenario.}

It is known that the coherent oscillations around the minimum of the potential
of a cosmic scalar background [1-4], such as the dilaton [5-7], put severe
cosmological bounds on the allowed value of the mass $m$ of that scalar field.
Such oscillations, however, become coherent, and then constrain the mass, only
when the possible spatial inhomogeneities of the background are negligible.

Spatial inhomogeneities of the dilaton background can arise either because of
thermal fluctuations, or because of the gravitational amplification of quantum
fluctuations. If the spatial inhomogeneities of thermal origin are diluted by
a subsequent inflationary expansion (or if they are absent simply because the
maximum temperature scale is lower than the Planckian temperature required for
thermal equilibrium), then their contribution to the energy
density (in a mode expansion of the background oscillations) becomes negligible
with respect to the mass contribution. The coherent oscillations of the
background,
with frequency $m$, can then dominate the dilaton energy density beginning at a
scale $H\simeq m$.

The same happens for the energy density stored in the quantum fluctuations, if
they are amplified with a spectrum which is flat or decreasing with frequency.
Also in that case the mass coherent contribution begins to dominate (with
respect to spatial inhomogeneities) the fluctuation
energy at $H \simeq m$, thus
avoiding a significant relaxation of the cosmological bounds, even for
negligible amplitude of the classical background oscillations [2].

If, on the contrary, the quantum fluctuations are amplified with a spectrum
which grows with frequency, their energy density stays dominated by the small
scale inhomogeneities until values of $H$ much lower than $m$. In that case the
bounds on $m$ can be alleviated, provided the classical background solution
approaches enough the minimum of the potential before the scale drops  to
$H=m$.

A growing spectral distribution of the dilaton fluctuations is a typical
outcome  of the
"Pre-big-bang" inflationary scenario [8,9], suggested by the duality properties
of the string cosmology equations [10]. In such context, as we shall see in
this paper, the relaxation of the bound has two important consequences: the
allowed range of dilaton masses becomes compatible a) with model of
supersymmetry breaking able to provide a natural resolution of the gauge
hierarchy problem [11], and b) with the possibility that our universe is
presently dominated by a relic background of dilaton dark matter. (It should be
mentioned that a growing scalar perturbation spectrum is also predicted by the
"hybrid inflation" model proposed by Linde [12] and recently generalized to the
class of "false vacuum inflation" [13] (see also [14]). Growing spectra,
moreover, have been shown to be necessary for a simultaneous fit of the COBE
anisotropies and of the observed bulk motion and large voids structures on a
$50 Mpc$ scale [15]).

In view of its importance, we start by recalling the standard arguments leading
to the bounds on the dilaton mass based on the coherent oscillations of the
classical background [1-7]. For scales $H<H_1\leq M_p$, where $M_p$ is the
Planck mass and $H_1$ the Hubble parameter at the time $t_1$
which marks the end of inflation (and which we
assume to coincide with the beginning of the radiation-dominated era), a
homogeneous dilaton background must satisfy the equation
$$
\ddot \phi +3H\dot \phi + {\pa V \over \pa \phi} =0 \eqno(1)
$$
(we neglect for the moment the friction term due to a possible dilaton decay,
with width $\Ga \simeq m^3/M_p^2$).
For values of $\phi$ sufficiently near to the minimum $\phi _0$ of $V$, we
approximate the potential by $V= m^2(\phi -\phi_0)^2/2$, with $m<H_1$. When
$H>>m$ one thus finds that $\phi$ approaches asymptotically a constant value
$\phi_1$,
$$
\phi =\phi_1 +\phi_2({H\over H_1})^{1/2} \eqno(2)
$$
so that, for $H<<H_1$, the distance from the minimum can be approximated by
$|\phi -\phi_0|\simeq |\phi_1 -\phi_0| = const$. Typically, for interactions of
gravitational strength, $\phi_0,\phi_1$ and $\phi_2$ are all of order $M_p$
but, without fine tuning, one expects that also $|\phi_1 -\phi_0|\sim M_p$. As
a consequence, when the scale $H$ drops to $H=m$, the homogeneous background
$\phi$ begins to oscillate coherently with frequency $m$ and initial amplitude
of order $M_p$, which decreases in time like $a^{-3/2}$ ($a$ is the scale
factor of the isotropic metric background). The associated energy density,
$\rho_\phi$, decreases like $a^{-3}$, starting from an initial value $\rho_i
\simeq m^2M_p^2$ which is of the same order as the radiation energy density
$\rho _\ga$ at that epoch. For $H<m$ the universe thus enters a phase of
dust-like matter domination, in which $\rho_\ga /\rho_\phi \sim a^{-1}$. This
leaves two possible alternatives open. If
$$
m \me H_2 \simeq 10^{-27} eV \eqno(3)
$$
where $H_2$ is the usual equilibrium scale corresponding to the
matter-radiation transition, then $\rho_\phi$ stays always smaller than the
critical density, and the oscillating dilaton background could survive until
tody. This possibility seems to be excluded by the present tests of the
equivalence principle, which imply [16,17]
$$
m \Me m_0 = 10^{-4} eV \eqno(4)
$$
(see however [18]). As a consequence, the dilaton must have decayed, i.e. the
energy stored in the coherent oscillations must have converted in radiation, at
a decay scale
$$
H_d = \Ga \simeq {m^3\over M_p^2} <m \eqno(5)
$$
However, since the
dilaton begins to dominate much before the nucleosynthesis scale
$H_N$, as
$$
H_N \simeq {(1 MeV)^2 \over M_p} << m_0 \me m , \eqno(6)
$$
one must impose that the reheating temperature $T_r$ associated to the dilaton
decay, such that
$$
T_r^4 \simeq M_p^2 H_d^2 , \eqno(7)
$$
be large enough to allow a subsequent nucleosynthesis phase. This means
$T_r \simeq (m^3/M_p)^{1/2} \Me 1 MeV$, namely
$$
m \Me 10^4~ GeV . \eqno(8)
$$
Moreover, the radiation temperature $T_d$ just before dilaton decay is
$$
T_d=T_m ({a_m\over a_d}) \simeq (mM_p)^{1/2} ({H_d \over m})^{2/3}
\simeq ({m^{11}\over M_p^5})^{1/6} \eqno(9)
$$
(the index $m$ means that the variable is to be evaluated at the scale $H=m$).
The reheating from $T_d$ to $T_r$ will thus produce an entropy increase
$$
\Da S = ({T_r \over T_d})^3 \simeq {M_p \over m} \eqno (10)
$$
In order to preserve any pre-existing baryon-antibaryon asymmetry, the
condition $\Da S \me 10^5$ should be satisfied [5-7]. Such a condition implies
$$
m\Me 10^{14}~ GeV \eqno(11)
$$
This last requirement could be alleviated, however, in the case of low-energy
(electro-weak, for instance) baryogenesis; in particular, in the case of
baryogenesis associated to the dilaton decay itself [7,11], occuring at scales
not much distant from nucleosynthesis.

These are the standard arguments, leading to bounds on $m$ which are
independent
of the inflation scale $H_1$, and which are crucially grounded on the
assumption that the asymptotic value $\phi_1$, approached by $\phi$ during its
evolution for $H>>m$, lies at a distance of order unity (in Planck units) from
the minimum $\phi_0$ of the potential.

One might thus be led to think that the bounds could be evaded if, owing to
some
mechanism, the initial amplitude of the classical background oscillations would
be lowered to $|\phi-\phi_0|<<M_p$. For instance, if the asymptotic value
$\phi_1$ would be fine-tuned to $\phi_0$, then $\phi - \phi_0$ would be no
longer constant but would become scale-dependent according to eq.(2). At a
scale $H$, the shift from the minimum would be typically of order $|\phi -
\phi_0| \sim M_p(H/H_1)^{1/2}$, thus leading to coherent oscillations for
$H\leq m$ with initial energy density $\rho_1 \simeq (m M_p)^2(m/H_1)
<(mM_p)^2$.

Even in that case, however, the scenario would not be free of problems. Indeed,
as pointed out in Ref.[2],
besides the classical oscillations one must
always take into account also the quantum fluctuations of the background,
$\da \phi = \chi$, amplificated by the inflationary evolution. The fluctuation
modes $\chi_k$ satisfy the equation
$$
\ddot \chi_k +3H\dot \chi_k +({k^2\over a^2}+m^2) \chi_k =0 \eqno(12)
$$
and for $m<H_1$ the spectral distribution of the associated energy density,
$\rho_\chi$, is
$$
k{d \rho_\chi \over dk} \simeq ({H_1 a_1 \over a})^4 ({k\over k_1})^{n-1}
\eqno(13)
$$
Here $k_1=a_1H_1$ is the maximum amplified comoving
frequency, and the spectral index,
$n=3-2\a$, is fixed by the power $\a$ determining the evolution (in conformal
time $\eta$) of the scale factor during inflation, $a\sim |\eta|^{-\a}$.

Let us discuss, first of all all, the case of the scale-invariant,
Harrison-Zeldovich spectrum ($n=1$), corresponding to de Sitter-like inflation
($\a=1$) with $H=H_1= const$. In this case all the modes $\chi_k$ contribute to
the energy density
$$
\rho_\chi (t) \simeq ({H_1 a_1 \over a})^4 = H_1^2 H^2 \eqno(14)
$$
with the same amplitude. (Note the condition $H_1<M_p$ to be satisfied in order
that $\rho_\chi$ does not overclose the universe during radiation dominance). A
generic mode $k>am$ begins to oscillate at a scale $H_k=k/a_k$, with an
amplitude
$\chi_k$ which is initially of order $H_1$, and which decreases in time as
$\chi_k \simeq H_1(H/H_k)^{1/2}$. When $H\simeq m$, the
non-relativistic modes ($k/a<m$) begin to
oscillate, with initial amplitude $H_1$ and frequency $\simeq m$,
and they immediately become dominant
with respect to the other modes, as their contribution to $\rho_\chi$ decreases
like $a^{-3}$ instead of $a^{-4}$. For $H\leq m$ we are thus in a situation
where, beside the energy density stored in the possible coherent oscillations
of
the classical background, $\rho_\phi = \rho_m(a_m/a)^3$, we must have,
necessarily, also some energy stored in the coherent oscillations of the
quantum fluctuations, with
$$
\rho_\chi \simeq m^2 H_1^2({a_m\over a})^3 \eqno(15)
$$

In this paper we want to discuss how the cosmological
bounds on the dilaton mass are relaxed when we move from a scenario in which
the fluctuation spectrum is flat or decreasing ($n\leq 1$), to another scenario
characterized by a growing ($n>1$) spectrum. We shall thus assume, henceforth,
that the classical oscillations (whose initial amplitude is model-dependent)
are always negligible, $\rho_\phi < \rho_\chi$, and that all
bounds on $m$ follow
from the cosmological amplification of the quantum fluctuations only. We will
obtain, in this way, the {\it maximum} (approximately
model-independent) allowed region in
parameter space.

The energy density (15) is smaller than the radiation energy $\rho_\ga$ when
$H=m$, but it grows with respect to $\rho_\ga$ as the curvature scale decreases
in time, until it equals $\rho_\ga$ at an initial scale $H_i$. If $H_i < H_2$,
which means
$$
m< H_2({ M_p\over H_1})^4 \eqno(16)
$$
(recall that $H_2$ denotes the usual matter-radiation transition scale of
eq.(3)), then $\rho_\chi$ stays always smaller than the critical density. If,
however, $H_i>H_2$ then
$$
H_i= m({H_1 \over M_p})^4 \eqno(17)
$$
and the dilaton must have already decayed, $H_d>H_0$, to avoid contradictions
with the presently observed matter density. This implies
$$
m \Me (M_p^2 H_0)^{1/3} , \eqno(18)
$$
where $H_0\sim 10^{-61}M_p$ is the present curvature scale. The dilaton decay
generates an entropy $\Da S= (T_r/T_d)^3$, where $T_r$ is the reheating
temperature (7) and $T_d$ the radiation temperature at the dilaton decay epoch,
namely
$$
T_d=T_i({a_i\over a_d})=(M_p H_i)^{1/2}({H_d\over H_i})^{2/3}
\simeq ({m^{11}\over M_pH_1^4})^{1/6} \eqno(19)
$$
This gives
$$
\Da S = {H_1^2\over m M_p} \eqno(20)
$$
If $m<10^4 GeV$ the reheating temperature is too low ($<1 MeV$) to allow
nucleosynthesis: we must impose that nucleosynthesis already occurred,
$H_i<H_N$, and that [5-7] $\Da S \me 10$, in order not to destroy all light
nuclei formed. If, on the contrary, $m>10^4 GeV$, the nucleosynthesis scale is
subsequent to dilaton decay, and the only
possible constraint is [5-7] $\Da S \me 10^5$ in order to preserve primordial
baryogenesis.

These are the conditions to be imposed if $m<H_1$. If $m>H_1=k_1/a_1$, then all
modes are always non-relativistic, and the spectral energy distribution becomes
[19]
$$
k{d \rho_\chi \over dk} \simeq H_1^4 ({ a_1 \over a})^3
({m\over H_1})^2 ({k\over k_1})^{n-1}    \eqno(21)
$$
This gives, for $n=1$,
$$
\rho_\chi(t) \simeq m^2 H_1^2 ({a_1\over a})^3 \eqno(22)
$$
The scale $H=H_1$ marks the beginning of coherent oscillations with frequency
$m$ and initial energy $\rho_i=m^2 H_1^2$ (hence $m< M_p$ to avoid overcritical
density). There are no further bounds on $m$, in this case, as the fluctuation
energy is dissipated before a possible dilaton dominance. Indeed, the scale
$H_i=H_1(m/M_p)^4$ corresponding to the equality $\rho_\chi =\rho_\ga$ is
always smaller than the decay scale, $H_d=m^3/M_p^2>H_i$.

The values of $(m,H_1)$ allowed by the previous constraints are shown in
{\bf Fig.1}. One can see that the bounds on $m$ are relaxed but, as stressed
in [2], too low values of $H_1$ are in general required to be compatible with
an interesting range of masses. The preferred supersymmetry breaking scale
$m\sim 1 TeV$, for instance, is forbidden unless $H_1\me 10^{-8} M_p$. Similar
values of $H_1$ ($\sim 10^{-6} - 10^{-9} M_p$) are required to be compatible
with the possibility of a not yet decayed, and presently dominating, dilaton.

The situation becomes even worse for quantum fluctuation with a decreasing
spectrum ($n<1$). In that case the initial amplitude $\chi_i$
of the coherent oscillations become larger, $\chi_i \simeq
H_1(H_1/m)^{(1-n)/4} >H_1$ and, as a consequence, a lower inflation scale $H_1$
 is required to be compatible with the same given value of $m$.

String cosmology, however, suggests a scenario in which the standard
radiation dominated era is preceeded by a so-called pre-big-bang phase,
describing the evolution from a flat, weakly coupled initial state [8]. The
universe superinflates, bends up and heats up to a maximum scale $H_1$, after
which curvature and temperature begin to decrease. The dilaton grows up to the
strong coupling regime, and its settlement to a constant value marks the
beginning of the standard cosmological evolution. The particle production
associated with the transition between pre and post-big-bang regime is
characterized by a growing spectrum [8,9], and imposes the constraint
$H_1<M_p$ to prevent an overcritical density of produced massless particles
(such as gravitons). In that context, the quantum fluctuations of the dilaton
background are also amplified with a growing spectrum, in particular with the
same spectrum of tensor perturbations
in the case of a vacuum, dilaton-driven
pre-big-bang [19]. (One may note that if such a spectrum would apply also to
the scalar part of the metric perturbations, it would impossible to explain the
anisotropy observed by COBE [20], which should then to be ascribed to other
sources. It should be stressed, however, that the dilaton perturbations and
the scalar perturbations of the metric background are not necessarily forced to
have the same spectrum, if other gravitational sources, beside the dilaton, are
present [19]).

Motivated by string cosmology, we shall thus discuss how the previous bounds on
the dilaton mass are to be changed if the fluctuation spectrum is growing with
frequency. If $n>1$ in eq.(13), the fluctuation energy (14) is initially
dominated by the maximum frequency mode $k_1= a_1H_1$, whose amplitude,
$\chi_1$, decreases in time as $\chi_1 \simeq (H H_1)^{1/2}$. As a
consequence, the coherent
oscillations with frequency $m$,
which begin at $H\simeq m$, may become dominant at a
scale $H_{nr}$ lower than in the previous case, such that $k_1/a_{nr}=m$,
namely
$$
H_{nr}= {m^2\over H_1} <m \eqno(23)
$$
Only for $H\me H_{nr}$ the contribution to $\rho_\chi$ of the small scale
inhomogeneities becomes negligible with respect to the mass contribution, and
the behaviour of $\rho_\chi$ becomes that of non-relativistic matter,
$$
\rho_\chi \simeq ({mH_1a_1\over k_1})^4 ({a_{nr}\over a})^3 =
m^4 ({a_{nr}\over a})^3 \eqno(24)
$$
corresponding to coherent oscillations with frequency $m$ and initial amplitude
$m$.

This energy density stays always smaller than the critical one if
$H_{nr}<H_2$, and also if $\rho_\chi$ equals $\rho_\ga$ at a scale $H_i<H_2$,
namely for
$$
m< (H_1H_2)^{1/2} ({M_p\over H_1})^2         \eqno(25)
$$
The same happens if $H_i>H_2$, where
$$
H_i={m^2\over M_p}({H_1 \over M_p})^3 \eqno(26)
$$
but the dilaton decays before becoming dominant, $H_d>H_i$. If on the contrary,
$H_i>H_2$ and $H_d<H_i$, the dilaton must have decayed (to avoid a present
overcritical density) when it was dominant and, as before, we are left with two
possible alternatives.

If $T_r<1 MeV$ we must impose that nucleosynthesis already occurred, i.e.
$H_N>H_i$ (with the scale $H_i$ of eq.(26)), and that $\Da S =(T_r/T_d)^3
\me 10$. Since
$$
T_d=T_i({a_i\over a_d}) \simeq (M_pH_i)^{1/2}({H_d\over H_i})^{2/3}=
({m^{10}\over M_pH_1^3})^{1/6} \eqno(27)
$$
the produced entropy to be bounded is, in this case,
$$
\Da S = ({H_1^3 \over m M_p^2})^{1/2} \eqno(28)
$$
The other possibility, $T_r >1 MeV$, allows a nucleosynthesis phase subsequent
to dilaton decay, so that the only bound is imposed by primordial baryogenesis,
$\Da S \me 10^5$. The case $m>H_1$, finally, provides the only bound $m<M_p$
exactly like in the previous case of a flat perturbation spectrum.

The allowed region in the plane ($m,H_1$), for the case of quantum fluctuations
with a growing spectrum, is illustrated in {\bf Fig.2}. The forbidden region
describes a sort of funnel wedged between the lower limit $m=m_0 = 10^{-4} eV$
allowed by the
equivalence principle, and the upper limit $m=M_p$ allowed by the closure
density. The bounds are relaxed with respect to the previous case, in such a
way that for $H_1\me 10^{-5} M_p$ practically all values of $m$ are allowed. In
particular, a dilaton mass in the $TeV$ range (required by models of
supersymmetry breaking [11]), becomes compatible with an inflation scale
$H_1 \simeq 10^{-5}M_p$, which seems to be the value suggested by the observed
COBE anisotropy [21]. Moreover, a universe presently dominated by a relic
background of not yet decayed dilatons becomes possible for
$$1~ eV \me m \me 100~ MeV \eqno(29) $$
and compatible with realistic (in a string cosmology context) inflation scales
$H_1 \geq 10^{-5} M_p$.

In conclusion, our qualitative analysis shows that the bounds on the dilaton
mass arising from the cosmological amplification of the quantum fluctuations
are less constraining when their spectrum is growing, like in a string
cosmology context, instead of being flat or decreasing like in other, more
conventional inflationary models. The bounds of {\bf Fig.2} define the
{\it maximum}
allowed region, corresponding to the case in which the classical dilaton
background seats exactly at the minimum of the effective potential, and
it is
neither oscillating nor running at the scale $H=m$. Such a region
would be of course reduced if classical background oscillations were to
be added to the quantum fluctuations.
It can be easily verified, however, that as
long as one can arrange a scenario in
which the distance of the background from the minimum decreases with the scale
like $H^{1/2}$ (or faster), the bounds on $m$ in the higher scales sector
$H_1 \Me 10^{-5}$ (which is the interesting sector for string cosmology) remain
practically unchanged with respect to the bounds reported in {\bf Fig.2}.

\vskip 2 cm
\noi
{\bf Acknowledgements} I wish to thank G.Veneziano for many helpful
discussions on the ideas reported in this paper, and for a critical
reading of the manuscript.

\vfill\eject

\centerline{\bf References}
\vskip 1 cm
\item{1.}G. D. Coughlan et al., Phys. Lett. B131 (1983) 59

\item{2.}A. S. Goncharov, A. D. Linde and M. I. Vysotsky, Phys. Lett. B147
(1993) 279

\item{3.}G. German and G. G. Ross, Phys. Lett. B172 (1986) 305

\item{4.}O. Bertolami, Phys. Lett. B209 (1988) 277

\item{5.}J. Ellis, D. V. Nanopoulos and M. Quiros, Phys. Lett. B174 (1986) 176

\item{6.}J. Ellis, N. C. Tsamish and M. Voloshin, Phys. Lett. B194 (1987) 291

\item{7.}B. de Carlos et al., Phys. Lett. B318 (1993) 447

\item{8.}M. Gasperini and G. Veneziano, Astropart. Phys. 1 (1993) 317;

Mod. Phys. Lett. A8 (1993) 3701

\item{9.}M. Gasperini and M. Giovannini, Phys. Rev. D47 (1993) 1529

\item{10.} G. Veneziano, Phys. Lett. B265 (1991) 287;

K. A. Meissner and G. Veneziano, Mod. Phys. Lett. A6 (1991) 1721;

A. A. Tseytlin, Mod. Phys. Lett. A6 (1991) 1721;

A. A. Tseytlin and C. Vafa, Nucl. Phys. B372 (1992) 443;

M. Gasperini and G. Veneziano, Phys. Lett. B277 (1992) 256

\item{11.}T. Banks, D. B. Kaplan and A. E. Nelson, UCDS/PTH 93-26, RU-37
(July 1993)

\item{12.}A. D. Linde, Phys. Lett. B259 (1991) 38;

Phys. Rev. D49 (1994) (in press)

\item{13.}E.J. Copeland et al., SUSSEX-AST 94/1-1, (astro-ph/9401011)

\item{14.}S. Mollerach, S. Matarrese and F. Lucchin, astro-ph/9309054

\item{15.}T. Piran et al., astro-ph/9305019

\item{16.}T. R. Taylor and G. Veneziano, Phys. Lett. B213 (1088) 450

\item{17.}J. Ellis et al., Phys. Lett. B228 (1989) 264

\item{18.}T. Damour and A. M. Polyakov, hep-th/9401069

\item{19.}M. Gasperini and G. Veneziano, Dilaton production in string cosmology
(to appear)

\item{20.}G. Smoot et al., Astrophys. J. 316 (1992) L1

\item{21.}M. White, Phys. Rev. D46 (1992) 4198;

T. Souradeep and V. Sahni, Mod. Phys. Lett. A7 (1992) 3541;

A. R. Liddle, SUSSEX-AST 93/7-2 (July 1993).

\vfill\eject

\centerline{\bf Figure captions}

\vskip 1 cm
\noi
{\bf Fig.1} Maximum allowed region (inside the full lines) for quantum
fluctuations amplified with a scale-invariant spectrum, and dominant with
respect to the classical background oscillations. The dilaton mass is given in
units of $m_0= 10^{-4} eV$. The lines marked by $a,b,c,d,e,f,g$ correspond
respectively to: a) $m=m_0$, lower bound on $m$ from the equivalence principle;
b) $H_1=M_p$, upper bound on $H_1$ from the closure density; c) $T_r=1 MeV$,
lower bound on the reheating temperature from nucleosynthesis; d) $m=M_p$,
upper
bound on $m>H_1$ from the closure density; e) $m=H_2(M_p/H_1)^4$, upper bound
on $m$ from the present matter-to-radiation energy density ratio; f)
$H_1^2/m M_p = 10^5$, upper limit on entropy production from primordial
baryogenesis; g) $H_1^2/m M_p = 10$, upper limit on entropy from
nucleosynthesis.

\vskip 1 cm
\noi
{\bf Fig.2} Maximum allowed region (inside the full lines) in the case of
quantum fluctuations amplified with a growing spectrum, and dominant with
respect to the classical background oscillations. The lines marked by $a,b,c,d$
are the same as in Fig.1. The other lines correspond to: e) $m=(H_1H_2)^{1/2}
(M_p/H_1)^2$, upper bound on $m$ from the present matter-to-radiation energy
density ratio; f) $(H_1^3/m M_p^2)^{1/2}=10^5$, upper limit on entropy
production from primordial baryogenesis; g) $(H_1^3/m M_p^2)^{1/2}=10$, upper
limit on entropy production from nucleosynthesis.
\end